\documentclass[structabstract,printer]{aa} 
\usepackage{graphicx}
\usepackage{natbib}
\usepackage{amsmath}
\usepackage{txfonts}
\usepackage{color}

\begin{document}

\title{Modes of clustered star formation}
\author{S. Pfalzner\inst{1} T. Kaczmarek\inst{1}, C. Olczak\inst{2,3,4}}
\institute{
\inst{1}Max-Planck-Institut f\"ur Radioastronomie, Auf dem H\"ugel 69, 53121 Bonn, Germany\\
\inst{2}Astronomisches Rechen-Institut (ARI), Zentrum f{\"u}r Astronomie Universit{\"a}t Heidelberg, M{\"o}nchhofstrasse 12-14, 69120 Heidelberg, Germany\\
\inst{3}Max-Planck-Institut f{\"u}r Astronomie (MPIA), K{\"o}nigstuhl 17, 69117 Heidelberg, Germany\\
\inst{4}National Astronomical Observatories of China, Chinese Academy of Sciences (NAOC/CAS), 20A Datun Lu, Chaoyang District, Beijing 100012, China\\
\email{spfalzner@mpifr-bonn.mpg.de}}
\date{ }

\abstract
   {The recent realization that most stars form in clusters, immediately raises the question of 
    whether star and planet formation are influenced by the cluster environment. The stellar density 
    in the most prevalent clusters is the key factor here. Whether dominant modes of clustered star formation exist is a fundamental question. Using near-neighbour 
    searches in young clusters Bressert et al. (2010) claim this not to be  the case. They conclude that - 
    at least in the solar neighbourhood - star formation is continuous 
    from isolated to densely clustered and that the environment plays a minor role in star and planet formation.}
   {We investigate under which conditions near-neighbour searches in young clusters can distinguish 
    between different modes of clustered star formation. }
   {Model star clusters with different memberships and density distributions are set up and near-neighbour 
    searches are performed. We investigate the influence of the combination of different cluster modes, 
    observational biases, and types of diagnostic on the results. 
   }
   {We find that the specific cluster density profile, the relative sample sizes, limitations in observations 
    and the choice of diagnostic method decides whether modelled modes of clustered star formation are detected 
    by near-neighbour searches.  For density distributions that are centrally concentrated  but span a wide density range (for example, King profiles) separate cluster modes are only detectable under ideal conditions 
    (sample selection, completeness) if the mean density of the individual clusters differs by at least a factor 
    of $\sim$65. Introducing a central cut-off 
can lead to underestimating
       the mean density by more than a factor of ten especially in high density regions. Similarly, the environmental effect on star and planet formation is underestimated for half of
       the population in dense systems.}
   {Local surface density distributions are a very useful tool for single cluster analysis, but only for high-resolution data. However, a
       simultaneous analysis of a sample of cluster environments involves effects of superposition that suppress characteristic features very
       efficiently and thus promotes erroneous conclusions. While multiple peaks in the distribution of the local surface density in star forming
       regions imply the existence of different modes of star formation, the reverse conclusion is
 {\em not} possible. Equally, a smooth distribution is {\em not} a proof of continuous star formation, because such a shape can easily hide modes of clustered star formation.}

\maketitle

\section{Introduction}

Most stars form in proximity to other stars within embedded clusters rather than being uniformly distributed throughout molecular clouds
\citep{1999A&A...342..515T,2000AJ....120.3139C,2003ARA&A..41...57L,2003AJ....126.1916P,2007prpl.conf..361A}. The density in young clusters in the
Milky Way varies over many orders of magnitude from $\ll$ 1 stars/pc$^3$ in relatively sparse clusters to $>$10$^5$ stars/pc$^3$ in the central areas
of dense clusters.  The key factor in determining the relative importance of the environment for star and planet formation is the stellar density in
the young clusters. Stars forming in the sparse cluster environments are largely unaffected by the presence of their fellow cluster members. By contrast, one can expect a strong influence on star and planet formation by the
environment in the densest of these young clusters. Theoretical investigations predict that this environmental influence on star formation might
manifest itself in a different initial mass function \citep{2006MNRAS.368..141F,2006ApJ...652L.129P,2012MNRAS.tmp.2706M}, the binary fraction
\citep{2011MNRAS.417.1684M,2011A&A...528A.144K} and the disc frequency in high stellar density environments
\citep{2001MNRAS.325..449S,2005A&A...437..967P,2006A&A...454..811P,2006ApJ...642.1140O,2010A&A...509A..63O}. Observations have found indications of a
dependence of these properties on the stellar density in young clusters \citep{1998ApJ...492..540H,2008ApJ...675.1319H,2010ApJ...718..810S}. 
In dense clusters interactions might lead to a lower disc frequency  resulting in a lower planetary system rate and 
different properties in the planetary system.

For the stellar population as a whole the question is whether the properties of prestellar cores largely determine the stellar properties as in
isolated star formation \citep{2004ApJ...601..930S,2005MNRAS.359..211L,2006ApJ...641L.121T} or whether most stars form in a more dynamic way, where
external forces and interactions dominate over initial conditions \citep[e.g.][]{2001MNRAS.323..785B,2006MNRAS.370..488B}.

So a fundamental question of current star formation research is whether there exists a type of cluster (stellar emembership, density) that is the dominant environment for
  star formation? At first sight it would seem easy enough to answer this by simply collecting cluster data and determine the distribution of
the mean density in young clusters. However, this is hindered by a number of obstacles. Most star formation occurs inside the spiral arms and close to
the center of the Milky Way where it is difficult to identify clusters due to our position  within the plane of the Galactic disc.
This means we have nothing like a complete census of the young clusters in the Milky Way. In principle, looking
at nearby galaxies should help, but the larger distance means that the detection of low-mass clusters is hindered by their low luminosity.

There are different strategies for tackling the issue indirectly. One way is to look at the initial mass function
\citep[e.g.][]{2007prpl.conf..149B,2008ApJ...681..771M,2009MNRAS.392.1363B,2012ApJ...748...14D} or the binary development
\citep[e.g.][]{1994A&A...286...84D,1998ApJ...499L..79B,1999A&A...341..547D,2008AJ....135.2526C,2009ApJ...707.1533F,2011A&A...528A.144K,2011MNRAS.417.1684M}
in different types of young clusters and compare them to the field properties. Similiarities 
i
are then interpreted as signs for a dominant cluster mode. However, since many cluster modes contribute simultaneously,
a one-to-one relation is difficult to establish.

Another method 
is to measure the local surface density distribution in different cluster
environments. Recently several observational studies \citep[e.g.][]{2009ApJS..184...18G,2010MNRAS.409L..54B,2012ApJ...745..131K} tried to answer above
questions by analyzing large samples of young stellar objects concerning their local surface density, $\Sigma$, predominantly in the solar
neighbourhood. Here it is argued that if different discrete modes existed they should manifest themselves as peaks in a surface density distribution
\citep[e.g.][]{1993ApJ...412..233S,2000AJ....120.3139C,2004MNRAS.350.1503W,2009ApJ...696...47W,2010MNRAS.409L..54B}.

This simple approach has the advantage that it does not rely on the definition of stellar groups, but
uses the local 
separation from the star to its nearest neighbours. The local surface density is simply defined as
%
  $ \Sigma = (n-1)/(\pi r_n^2)$
%
where $n$ is the considered number of nearest neighbours including the star itself and $r_n$ is the distance to the $n$-th neighbour.
Higher values of $n$ give a lower spatial resolution, but smaller fractional uncertainty
\citep{1985ApJ...298...80C,2009ApJS..184...18G}.

Using this method \citet{2010MNRAS.409L..54B} found no peaks in the combined surface density distribution of several clusters in the solar
neighbourhood (see their Fig.~1). They concluded from the absence of such peaks that star formation is continuous from isolated to densely
clustered. In addition, they deduce a mean stellar surface density of 20~stars/pc$^2$ for the star forming regions in the solar neighbourhood and
concluded that the environment plays a minor role in star and planet formation because only a small fraction of stars is found in high-density
regions.

In the present study we will discuss the effect of different cluster density profiles, the dependency on the sample selection and the influence of
observational constraints on the obtained results. We will demonstrate that local surface density measurements are rather limited in their ability to
determine different star formation modes due to superposition effects.  Therefore the question whether dominant modes of clustered star formation
exist in the solar neighbourhood is still open.

\section{Method}
\subsection{Cluster types}
\label{sec:cluster-types}

The determination of the general shape of the stellar density distribution of young clusters can be observationally challenging. Due to the presence
of a significant amount of dust in young embedded clusters, not all stars are yet visible and even in young exposed clusters crowding in the
central high density regions poses problems even with high resolution instruments like the HST \citep[e.g.][]{1994AJ....108.1382M}.

A century ago \citet{1911MNRAS..71..460P} found that 
%
\begin{equation}
  \label{eq:setup__plummer__density}
  \rho_P(r) = \left(\frac{3M}{4\pi a^3}\right)\left(1 +\frac{r^2}{a^2}\right)^{-5/2}, 
\end{equation}
provides a good fit to the density distribution of 
globular clusters. Here $M$ is the total cluster mass and $a$ is
the Plummer radius, a scale parameter for the cluster core size $r_c$ 
This model is widely used for all types of star clusters, largely thanks to its success in fitting globular cluster profiles, but also
because of its convienient analytical form.

\citep{1966AJ.....71...64K}
found an improved empirical law,
%
%
leading to the so-called family of King models. These consist of an energy distribution function of
the form
\begin{equation}
  \label{eq:setup__king__distribution_function}
  f_{K} ( \mathcal{E}) = \left\{ 
    \begin{array}{l@{\quad:\quad}l} 
      \rho_{1}(2\pi \sigma_K^{2})^{-3/2} ( e^{\mathcal{E}/\sigma_K^{2} } - 1 ) & \mathcal{E} > 0 \\ 
      0 & \mathcal{E} \leq 0
    \end{array}\right.,
\end{equation}
with $\mathcal{E} = \Psi - \frac{1}{2}v^2$ and $\Psi = -\Phi + \Phi_{0}$ being the relative energy and relative potential of a particle,
respectively. Here $f(\mathcal{E}) > 0$ for $\mathcal{E} > 0$ and $\sigma_K$ is the King velocity dispersion. 
The stellar density distribution can only be obtained by numerical integration. 
The
King paramter $W_0 = \Psi/\sigma_K^{2}$ characterizes the sequence of King profiles with decreasing relative size of the cluster core
$r_c/r_{\text{hm}}$ for increasing $W_0$, where $r_{\text{hm}}$ is the half-mass radius. 

In the following we investigate two types of model clusters - those based on Plummer and King distributions. While the Plummer distribution is well
approximated by a King model with $W_0 \approx 4$, young clusters are best represented by King models with $W_0 \ge 7$
\citep[e.g.][]{1998ApJ...492..540H,2004AJ....127.1014S,2010MNRAS.409..628H}. Thus the term ``King model'' is used here as equivalent to King
distributions with high $W_0$.

%





\subsection{Diagnostics}

\begin{figure*}
  \centering
  \includegraphics[width=17cm]{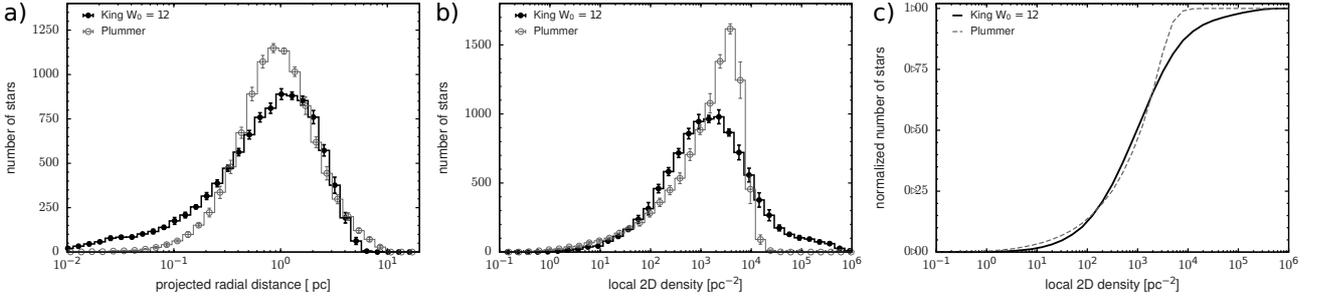}
  \caption{Comparison of  model clusters with 10 000 stars and a half-mass radius of 1 pc obeying a Plummer distribution (open circles) w
ith those of a King ($W_0$ = 12). Here a) shows the number of stars as a function of the radial distance to the cluster center, whereas b) shows the number of stars of a given surface density and c) the same in cumulative normalized form.}
\label{fig:comparison_number_vs_proj_dist_and_number_vs_local_2D}
\end{figure*}

In order to determine the conditions for which local surface density allows to distinguish different spatial modes of star formation, we construct a representative set of numerical cluster models that spans the expected parameter space. We generated model clusters with Plummer and King stellar density profiles containing
100, 1000 and 10000 stars.  Each cluster has a half-mass radius of $r_\text{hm} = 1$\,pc. So configurations with different numbers of
stars imply different volume und surface densities. In our model clusters the mean surface densities are 12.6, 126 and 1260 stars/pc$^2$,
respectively. The distributions have been set up as single stars only, so without primordial binary population.

To ensure equally statistically significant results each cluster population was generated repeatedly with different random seeds for a total of $10^5$ stars for each of the considered cases. 

We used a tree-based algorithm to reduce the computational effort for the near neighbour search \citep{2004physics...8067K}.

As already pointed out by \citet{1985ApJ...298...80C} an intermediate number of neighbours 
has the  advantage of neither missing small dense structures nor introducing artificial overdensities produced by strongly bound multiple systems.
We tested the influence of 
the number of nearest neighbours (3-27) on the resulting surface density diagnosed for our
King models. For the clusters with 1000 and 10 000 stars no obvious difference was visible in the results averaged over the set of simulations.  Only the results for the cluster consisting of 100 stars
  depends slightly on the number of neighbours considered.  However, 
even these differences 
are within the error bars. So we included the contribution of 8 nearest neighbours throughout our investigation.


\subsection{Model vs. observed clusters}

\label{sec:comparison-with-real-clusters}

In some respect our model clusters represent the ideal of what one would like to observe. However, 
In observations of even the closest star forming regions it is nearly impossible to detect
each and every star of the cluster. One reason is that
due to limitations in the spatial resolution of telescopes, crowding becomes a severe problem in the central regions of dense clusters. For example,
the Spitzer Space Telescope as used in the study of \citet{2010MNRAS.409L..54B} can only marginally resolve the inner 0.3\,pc of the Orion Nebula
Cluster. To avoid observational biases due to crowding they excluded this inner cluster area from their analysis.  This means that so-obtained values of the average stellar density 
only regarded as lower limits. 
For high-resolution telescopes like the HST
this is less of a problem.

Another limitation is the maximum contrast an instrument can image. 
This means that low-mass stars are less likely to be detected close to massive stars and therefore the 
surface density around massive stars, which are mostly located in the central dense area is underestimated.

Finally, magnitude limits of a given survey impose a limit on the faintest observable isolated object. With decreasing mass the number of stars in a
star cluster grows rapidly, so the estimated density is a strong function of the
magnitude limit. Usually, field contamination 
imposes another
serious observational bias. However, the young members of star forming regions can usually be rather well separated from the much older population of
field stars.

The observational limitations outlined above basically affect studies of any star-forming region. The effect of all these
  limitations is to lower estimates of the cluster density. This is particularly true for the maximum local density that is typically highest where
  crowding and massive stars impose the most severe observational biases.

\section{Single clusters}
\label{sec:single-clusters}

\subsection{Cluster density profile}


\begin{figure*}
  \centering
  \includegraphics[width=\linewidth]{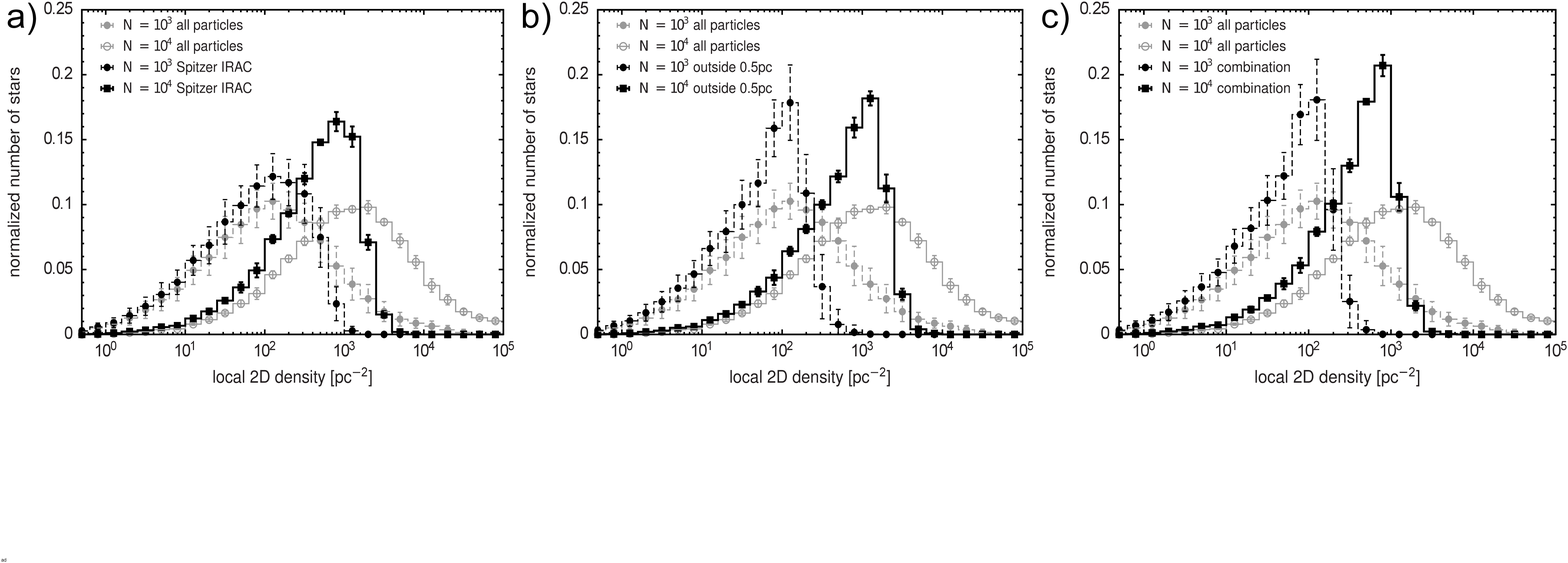}
  \caption{Comparison of the effect of different observational biases (black) on the unbiased local 2D density distribution (grey) for star clusters
    consisting of 1k (dashed) and 10k (solid) particles with a King parameter of $W_0 = 12$. \textbf{a)} Resolution limit of the IRAC camera of the
    Spitzer Space Telescope. \textbf{b)} Cutout of the densest inner 0.5\,pc. \textbf{c)} Combined effects a) and b).}
\label{fig:comparsion_multiple_filters}
\end{figure*}

First we compare single model clusters with a Plummer profile to those with a King ($W_0 = 12$) profile. Note that all models have normalized
half-mass radii ($r_\mathrm{hm} = 1$\,pc).  Plummer models are commonly used as initial models for numerical simulations of young star
clusters. However, observations of very young star clusters ($<$ 3Myr) typically show a more concentrated distribution close to that of an isothermal
sphere. From a numerical point of view a King model with $W_0 = 12$ is a rather good representation of such an
isothermal sphere. The basic difference between these two models is 
the stellar
density in a King model with $W_0 = 12$ increases much more towards the cluster center than in a Plummer model.

This is clearly visible in Fig.~\ref{fig:comparison_number_vs_proj_dist_and_number_vs_local_2D}a) which shows the projected radial number profile of
both models. Their maxima roughly correspond to the half-mass radius $r_\mathrm{hm}$. Whereas the Plummer-model clusters (gray lines) contain only a
small fraction of stars at small (projected) distances to the cluster centre, their fraction in King-model clusters (black lines) is considerably larger. 
In the local surface
density plot (Fig.~\ref{fig:comparison_number_vs_proj_dist_and_number_vs_local_2D}b) this translates into a sharply peaked asymmetric distribution for
the Plummer model and a Gaussian-shaped distribution for the King model. Most stars in the Plummer-shaped cluster share the same local density that
marks roughly the maximum local density of the entire cluster. In contrast, the King-shaped cluster has a long high-density tail that extends well
beyond the maximum local density of the Plummer model. In the cumulative local surface density distribution
(Fig.~\ref{fig:comparison_number_vs_proj_dist_and_number_vs_local_2D}c) this difference is encoded in the steeper slope at the end of the distribution
for Plummer-type clusters.





\subsection{Incompleteness}
\label{sec:compensation-for-incompleteness}

As described in Sec.~\ref{sec:comparison-with-real-clusters}, observations of real star clusters always suffer from observational limitations and
potentially influence the resulting surface density distribution. Here we mimic these observational limitations by applying ``filters'' to the data in our diagnostics. First we emulate the observational resolution of the Spitzer's IRAC camera of $2.5''$ for a cluster at the same
distance as the ONC corresponding to a resolution of 1035 AU $\approx$ 0.005pc. 
In our diagnostics we scan all particles and mark those which lie in projection within 1000\,AU from the current star as not being observable.

Fig.~\ref{fig:comparsion_multiple_filters}a) demonstrates the effects of this observational limitation for the King model cluster with $N=10^3$ and
$N=10^4$ stars, where grey indicates the case without filter and black the filtered case.  
Observational limitations lead to the neglect of any stars with local surface densities
exceeding roughly ten times the median density of stellar system. In the intermediate density regime a slight increase of the counted number of stars
is seen, because in high density areas the local surface density is reduced around those stars remaining observable. Here and in the following the number of stars in the
distributions have been normalized to the total number of stars in the sample.
 
So adopting the Spitzer-like resolution significantly reduces the number of stars at the high-density end. The average number of observed stars in the
filtered case reduces for the cluster containing 100 stars to 99, that with 1000 stars to 864 and the one with 10 000 to about 6460. In addition, for
the dense clusters the limited resolution renders the existing Gaussian-like shape as a much more peaked curve with a steep decline at high densities
- very similar to a Plummer distribution (cf. Fig.~\ref{fig:comparison_number_vs_proj_dist_and_number_vs_local_2D}).  As a consequence of the
observational limitation the median observed density of the densest of our model cluster would be reduced to less than half its real value (see
Table~\ref{table:nacc2}).



The problem of crowding is often circumvented by excluding stars in central high stellar density regions from the sample. Here we mimic this by
excluding all stars closer than $0.5pc$ to the cluster centre (Fig.~\ref{fig:comparsion_multiple_filters}b).
As in the case for the Spitzer resolution limitation, the relative number of stars with low local 2D densities ($\lesssim 10^2$pc$^{-2}$) remains
nearly unchanged, the number of stars with high local 2D densities (larger then ten times the median density of the cluster) is entirely removed,
while the number of stars with intermediate local 2D densities rises by adopting this filter. However, this time the increase of the number of stars
with intermediate densities is much more pronounced than in the Spitzer resolution case because of boundary effects at the filter cutoff radius. 
The result is an apparent increased number of stars with intermediate densities. 

Combining both (Fig.~\ref{fig:comparsion_multiple_filters}c), observational limitations lead to a significant underrepresention of the number of stars
in high-density regions of dense clusters. As a result an observer would underestimate the median local density by more than a factor of two and the
average local density by more than an order of magnitude (see Table~\ref{table:nacc2}). The quoted values can only be regarded as lower limits.


\section{Multiple modes}
\label{sec:multiple-modes}

In the following we analyse idealized samples of stars constructed from different cluster modes. Technically this is achieved by scaling the data sets
from Section~\ref{sec:single-clusters} accordingly. The aim is to determine under which circumstances one would be able to detect different cluster
modes from the (cumulative) surface density distributions.

\begin{figure}
 \centering
  \includegraphics[width=\linewidth]{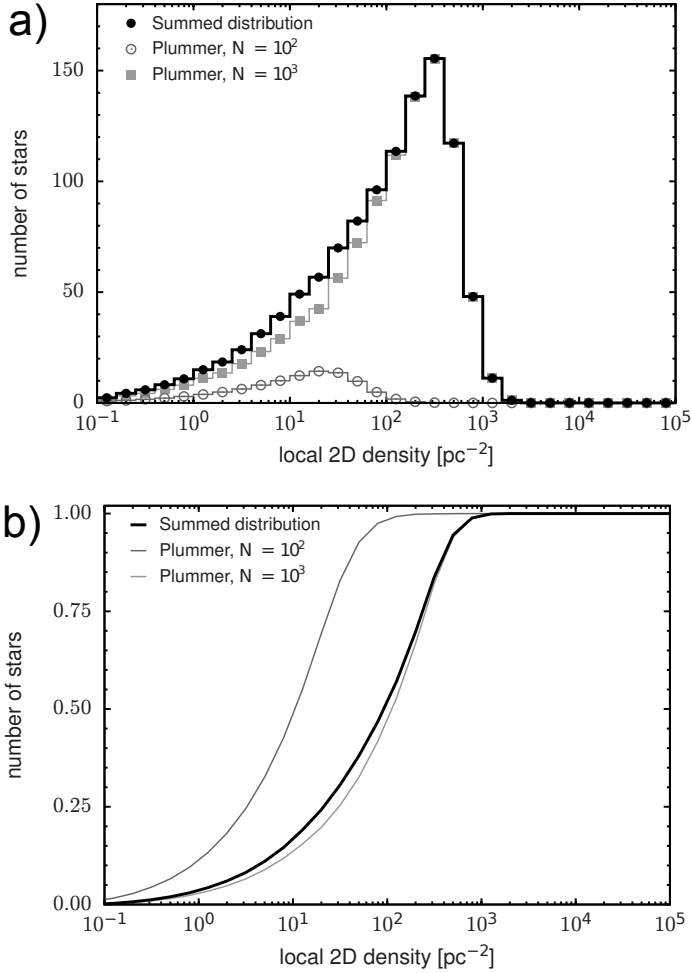}
  \caption{a) Differential and b) cumulative local surface density distribution for stars from two Plummer-shaped model clusters of approximate mean
    density of 12.6\,pc$^{-3}$ and 1260\,pc$^{-3}$, where the first one contains 100 stars and the latter 1000 stars.}
\label{fig:2Plummer_100_1000}
\end{figure}

\subsection{Relative sample size}

In reality sample sizes from different clusters often differ considerably. In many cases only a few tens of data points are available for low-mass
clusters but several hundreds to thousand for high-mass clusters like the ONC. Therefore, high-mass clusters might dominate the results. In order to
test the limitations we start with a model consisting of two modes of clustered star formation - one compromising a denser and the other a less dense
environment.

Combining two Plummer-type clusters, where one has a ten times higher median density than the other, Fig.~\ref{fig:2Plummer_100_1000}a) shows the differential local surface density distribution  and 
Fig.~\ref{fig:2Plummer_100_1000}b) its cumulative form for the case where one cluster corresponds to
the 100 and the other to the 1000 star models described in section~\ref{sec:single-clusters}. This illustrates a situation where 10 times as many
stars formed in the denser environment than in the less dense one. It can be seen that one would not detect two peaks. Similarily the
cumulative surface density distribution increases steadily and does not show any "bumps" although two different cluster modes were present.

This demonstrates, that the actual sample-size can mask an existing bi-modal clustered star formation process. We tested the
maximum possible difference in sample size that allows the identification of existing cluster modes and find that generally the sample sizes must not differ by less than a factor of five for existing cluster modes to be identifiable.

In reality, one either considers a smooth distribution of young stars throughout a single cloud or one combines the results from multiple distinct
clusters. It is obvious that above reservations apply in the first case. In the second case one could argue that there will be many more stars in
high-mass clusters than in low-mass clusters, but many more low-mass clusters than high-mass clusters. So in principle one can construct equal-sized
samples.


\subsection{Two equal modes of clustered star formation}

\label{sec:two-equal-modes}

In this section we analyse the case of an idealized sample combining two identical sample sizes of two different cluster modes. So we treat the case
where stars form with equal likelihood in one of two clustered modes.

\begin{figure*}
  \centering
  \includegraphics[width=17cm]{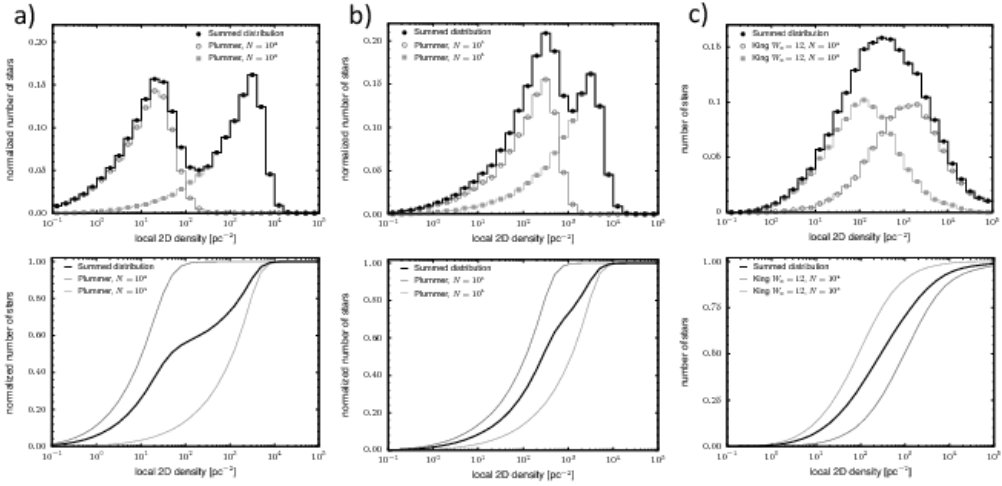}
  \caption{Differential and cumulative local surface density distribution for stars from two Plummer-shaped (a and b) and two King-shaped (c) model
    clusters. The approximate mean densities of each individual cluster is 10 stars pc$^{-3}$ and 1000 stars pc$^{-3}$ in a) and 100 stars pc$^{-3}$
    and 1000 stars pc$^{-3}$ in b) and c). For each cluster mode 10~000 stars were considered.}
\label{fig:two_modes}
\end{figure*}

We start with two Plummer-type clusters, where one has a hundred times higher median density than the other. Fig.~\ref{fig:two_modes} a) shows the
surface density on top and its cumulative form underneath for the case where one cluster corresponds to the 100 and the other to the 10~000 star models.
Scaling is applied to avoid non-detection of cluster modes due to sample size effects.  

The combined surface density shows a strong double peak and the cumulative distribution a saddle point. These features are still visible if the
density in one cluster is only 10 times that of the other cluster (see Fig.~\ref{fig:two_modes}b).

These multiple peaks in the surface density distribution and the "bumpy" nature of the cumulative distribution is what is expected for multi-modal
clustered star formation.  Conversely, the {\em absence} of these features is often taken as proof of continuous star formation ranging from low to
high density regions \citep[see, for example,][]{2010MNRAS.409L..54B}. It is argued that the peaks are so densely packed that the result is a
continuous function.  We will show that this argument is only valid under very specific conditions which are usually not fulfilled in young cluster
environments.

%

As mentioned above Plummer profiles are widely used in theoretical investigations due to the existence of an analytical solution. However, they seem less suitable for 
modelling young clusters. King profiles with high $W_0$ 
are regarded as a better choice. So performing the same investigation as above but now for two of
the King-type clusters 
deviating by a factor of 10 in density, we obtained results quite
different from the Plummer case. Instead of two peaks a single one appears in the surface density distribution (Fig.~\ref{fig:two_modes}c) which is no
longer "M"-shaped but nearly Gaussian and wider than in case of a single King-shaped cluster.

As a result the cumulative surface distribution (Fig.~\ref{fig:two_modes}c) looks very much like that of a single cluster with only a slightly
different slope. So despite being the result of two distinct modes of clustered star formation with a factor of 10 difference in mean cluster density,
this fact would neither be inferred from the differential nor the cumulative local surface density distribution in this case.

The reason that the two different modes of clustered star formation are detectable for Plummer-type but not for King-type clusters is the
different shape of the surface distribution of each individual cluster at the high-density end. For King-type clusters the high-density tail of the
lower-density cluster overlaps with the low-density end of the high-density cluster, creating a peak in the middle between the two mean cluster
densities. As there is no high-density tail in Plummer-type clusters
the steep drop leads to two clearly distinct
peaks. Consequently, the threshold for identifying distinct peaks in superpositions of King-type cluster modes is much higher and requires a
  ratio of the median densities of~$\sim$65.

The result that the shape of the distribution is relevant for the detectability of different cluster modes, does not only hold for the cases of Plummer and King models, but
applies to other distributions as well: distinct cluster modes are easily detectable for
narrow distribution whereas concentrated but broader distribtions can hide such modes. In the following we will continue to speak of King-type clusters, but the reader should keep in mind that this is valid for any type of broader distribution.

For concentrated King-type clusters
the absense of peaks in the surface density distribution
therefore neither allows the conclusion that there are not multiple modes of star formation present nor that star formation is continuous over all
cluster densities. At the same time a smooth surface distribution does {\em not} allow one to draw the conclusion that no distinct scale for YSO
clustering within nearby star-forming regions exists as, for example, recently stated by \citet{2010MNRAS.409L..54B}.

In view of the above findings 
local surface density distributions are limited in their informative value concerning existing modes of clustered star formation.

\subsection{Observational biases}

\begin{figure}
  \centering
  \includegraphics[width=\linewidth]{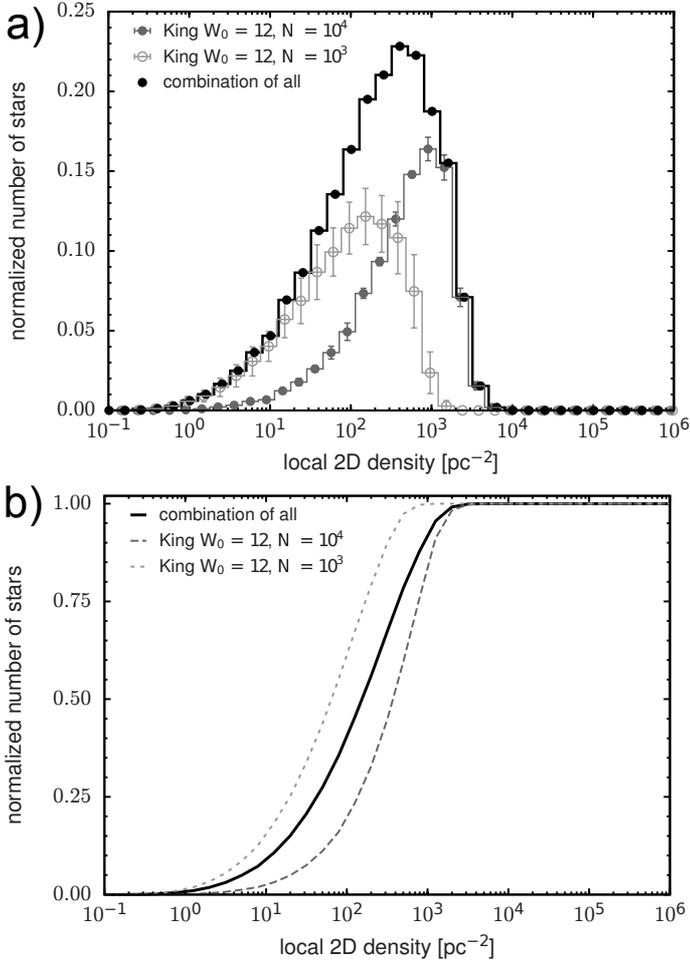}
  \caption{Same as Fig.~\ref{fig:two_modes}c but this time observational limitations are modelled for two clusters with a King profile (see
    Fig.~\ref{fig:comparsion_multiple_filters}). }
\label{fig:combined_observed}
\end{figure}

We just showed that for two King-type clusters, which differ in density by a factor 10, only a single peak appears in
the surface density profile. How far do observational limitations 
affect above
result?  Fig.~\ref{fig:combined_observed} shows the surface density distribution that results from two observationally limited King model clusters as
shown in Fig.~\ref{fig:comparsion_multiple_filters}. It can be seen that observational limitations lead to a non-physical cut-off at the high-density
end of the surface density distribution. Although the observational limitations lead to an under-representation of high-density areas, the two
underlying cluster modes are still not detected. Different cluster modes are only revealed if the peak densities of the two modes differ by more than a
factor $\sim$ 65.

\subsection{Multiple modes of clustered star formation}

\begin{figure}
  \centering

  \includegraphics[width=\linewidth]{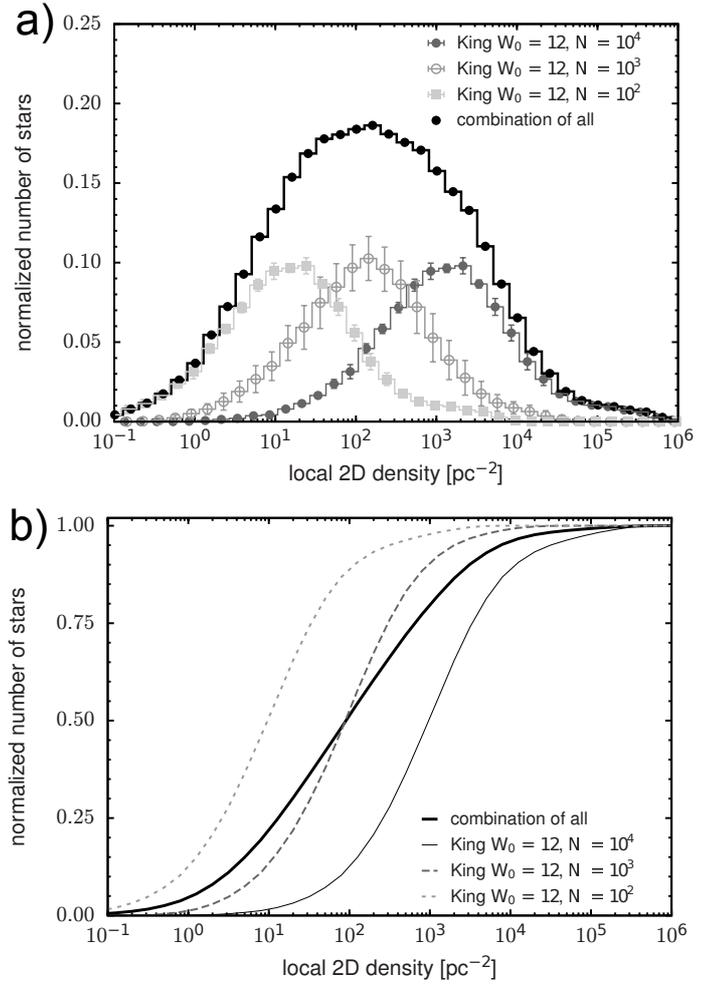}
  \caption{Local surface density distribution of the superposition of three King-type cluster modes with $10^2$, $10^3$, and $10^4$ stars,
      each with the same total stellar population size of $10^5$.}
  \label{fig:comination_three_clusters}
\end{figure}

If there are more than two modes of clustered star formation, the surface density distribution as diagnostic of multiple modes becomes increasingly
unreliable. Fig.~\ref{fig:comination_three_clusters} shows the combination of three model King-type clusters (non-detection limited cluster A, B and C
in Table~\ref{table:nacc2}) of different average density but with an equal number of stars in each mode. As in the case of two cluster modes, here
again the underlying three cluster modes would not show up as separate peaks 
but one obtains a more or less Gaussian-shaped
smooth distribution with a single (although this time broader) peak. In the cumulative surface plot this is represented by a smooth but somewhat flatter curve than the ones for the single clusters. This might possibly open up a way to detect the underlying cluster
modes.

We want to emphasize that we do  not advocate that all star formation happens in two, three or more modes but that surface density distributions are of limited use in inferring underlying modes of clustered star formation. Especially in the solar neighbourhood there are so far no indications for different modes of star formation. However, on Galactic scales that might, at least for massive clusters, be different \citep{1998mcdg.proc...37H, 2001ApJ...563..151M,2009A&A...498L..37P}.

\section{Influence on star and planetary system formation}

Observations often apply the technique of surface density plots to find out to 
what degree the cluster environment influences planet and star formation. These studies presume a density limit above which they assume that the
interactions between the stars become important.  Determining the relative proportion of stars that reside in areas with stellar densities above and
below that limit, this is then used as argument for or against the importance of the environment for star and planet formation.

Often the value of 10$^4$ stars pc$^{-3}$ \citep[see][]{2005ApJ...632..397G} is quoted as threshold for the cluster environment playing a role or
not. \citet{2009ApJS..184...18G} translated this into a local surface density exceeding 200 star pc$^{-2}$ \citep[see][]{2010MNRAS.409L..54B}. These
values are just rough estimates 
and it should be kept in mind that this value of the local surface density limit at which environmental effects play a role depends strongly on the actual aspect of star and planet
formation one considers. Stellar mergers, disc destruction or modifications of the disc structure
will correspond to very different local surface density limit.


\begin{table}
\begin{center}
\begin{tabular}{r *{5}{c}}
Property & Cluster A & Cluster B & Cluster C  \\[0.5ex]
\hline
\\[-2ex]
\multicolumn{3}{l}{Model clusters} \\
\hline
No. stars      & 10 000       & 10 000        & 10 000   \\[0.5ex]
median density & 12.6   & 126  & 1260  \\[0.5ex]
average density   & 21.2 & 530  & 11000 \\[0.5ex]
above threshold & 0        & 3300     & 8000 \\[0.5ex]
\hline
\\[-2ex]
\multicolumn{3}{l}{Spitzer resolution sample}\\
\hline
\\[-2ex]
No. stars      & 9923         & 8638         & 6458 \\[0.5ex]
median density & 12.6& 79.4 & 501  \\[0.5ex]
average density   & 21.2 & 151 & 668 \\[0.5ex]
above threshold& 0          & 1813        & 4327    \\[0.5ex]
\hline
\\[-2ex] 
\multicolumn{3}{l}{Radial cut-off and Spitzer resolution sample}\\
\hline
\\[-2ex]
No. stars      & 8476         & 7246         & 5990 \\[0.5ex]
median density & 7.9 & 50.1 & 501   \\[0.5ex]
average density   & 11.7 & 77.1 & 518  \\[0.5ex]
above threshold       & 0  & 290          & 3893 \\[0.5ex]  

\end{tabular}
\caption{Properties of the King-type cluster models used in Section~\ref{sec:multiple-modes}. The "above the threshold"-lines denote the "detectable" number of stars in an environment where the stellar density exceeds 200 stars pc$^{-2}$. The densities are median and average {\em surface} densities and are given in units of pc$^{-2}$.
  \label{table:nacc2}}

\end{center}
\end{table}

For the moment we take the estimated local surface density threshold -- 200 stars pc$^{−2} $ -- at face value to investigate how the cluster profile,
sample and incompleteness effects influence the estimate of the relative importance of the cluster environment on star and planet formation.  Returning
to Fig.~\ref{fig:comparsion_multiple_filters the effects of observational limitations 
lead to underestimating high local surface densities for the limits applied in Bressert et al. (2010).} For the three cluster modes
of different densities Table~\ref{table:nacc2} provides -- in dependence of observational limitations -- the number of observed stars, the resulting
change in average and median surface density, and the number of stars detectable above the local surface density threshold of 200\,stars\,pc$^{-2}$.

The values in our model clusters are scaled in such a way that all
three clusters contain 10~000 stars but their densities differ by a factor 10 and 100, respectively. In the least dense cluster no stars are located
in regions above the local surface density threshold, whereas 80\,\% of stars in the densest cluster encounter higher local densities and are thus
potentially effected 
by the cluster environment.

For a Spitzer resolution-limited sample 
obviously the densest cluster has the
largest number of undetected stars in high-density regions. However, in relative terms it is the same in intermediate- and high- density clusters - in
both cases observational limitations result in missing $\sim$ 45\% of the stars potentially affected by the environment.

Excluding the central area of the cluster from the study (see Bressert et al. 2010 ) again lowers the number of detected stars in high-density regions. If such a cut-off is applied in our high-density and
even intermediate density clusters the mean and average surface density are underestimated. However, whereas in high-density clusters resolution limitations already eliminate most of the stars affected by the high stellar density, in intermediate density clusters stars that would normally be resolved are heavily affected by excluding the central area. Bressert et al. state that excluding the central areas would at most effect the number of effected stars by a factor of 2. For our model cluster~B the cut-off
procedure and the Spitzer limitations would reduce the number of stars above the threshold to less than a tenth of its real value.

\section{Summary and Conclusions}

We investigated under which circumstances categorical distributions of local surface densities of young stellar objects -- here shortly referred
to as \emph{surface density plots} -- are suitable tools for investigating modes of clustered star formation and the dynamical influence of the star
cluster environment on star and planet formation. Using different types of model star clusters we demonstrate how sensitive the results depend on the
actual cluster density profile. Whereas for  narrow (for example Plummer-shaped) density distributions discrete cluster modes are easily
identified as multiple peaks in the surface density plot; this is often not the case for distributions that span over a wider density range - for, example, concentrated King-type density distributions.
 Our findings imply that surface density plots of star-forming regions will not show multiple peaks
unless the median density of the individual cluster modes differs by more than a factor of $\sim$65.

 The relative population size plays as well a role. Only if they do not differ by more than a factor of 5 the detection of discrete modes in the surface density plot is
possible.  
Even, if one constructs equal sized samples, there might arise difficulties. If one combines different low-mass clusters to a single sample, it is
very difficult to garantuee that they are in the same evolutionary stage. The cluster age is at least for embedded clusters not a reliable indicator
for their dynamical stage.  The reason is that if star formation is ongoing and accelerated then averaging will always lead to approximately the same
mean cluster age of $\sim$1-2Myr. So a cluster just starting to form stars and one that has nearly finished the star formation process will both be
attributed the same age. However, during that phase the cluster size, profile and surface density evolves considerably
\citep{2009A&A...498L..37P,2011A&A...536A..90P,2012Parmentier}. Including such different clusters in the same sample would lead to erroneous
results. This means that the right sample choice is vital to determine whether a dominant mode of clustered star formation exists.

 This means  that although one can conclude from multiple peaks in the surface density plot on the existance of
discrete modes of clustered star formation, the reverse is not possible. We point out that unlike assumed in recent publications
\citep[e.g.][]{2010MNRAS.409L..54B} a smooth surface density plot does {\em not} rule out the existence of dominant modes of clustered star
formation. We thus caution against the use of surface density plots to determine whether dominant modes of clustered star formation exist.

However, surface density plots are potentially very useful in determining the dynamical influence of the cluster environment on star and planet
formation. Yet a robust estimate requires high-resolution observations of rich star clusters to map the entire stellar population. Here we 
demonstrated that 
excluding regions of high local surface density in rich star clusters (like in Bressert et al. 2010)
leads to underestimating the average local surface density {\em not} as estimated in their study by at most a factor of two but by up to more than an order of magnitude.  Observations with instruments other than Spitzer (such as HST) are important for determining high surface density regions in  such clusters.

Another limitation that biases our understanding of star formation modes arises from restrictions of observational samples to the solar
neighbourhood. 
Although
there are good reasons for this approach such as sample completeness, one has to be aware that these results cannot be generalized to the
Galaxy as
for example, starburst clusters with their mostly much higher local surface densities
are excluded.

Similarly, the age of the clusters included in the sample is an important factor. Dynamical interactions and stellar evolution in star clusters induce
cluster expansion and hence act to lower their median local surface density with time.  This effect becomes even more pronounced if gas expulsion is
taken into account \citep[see e.g. the review of][]{2010RSPTA.368..829V}. Hence using a sample with a spread of cluster ages leads to an underestimate
of the median local surface density of a given mode. This is of particular relevance for low-mass clusters that are expected to dissolve faster due to
their short relaxation time.


In summary, a consistent analysis of the modes of clustered star formation requires a sample of isochronal clusters unlimited in mass and the
development of a tool suitable to reveal potential discrete modes.

\begin{acknowledgements}We would like to thank the referee for the very constructive comments and A. Stolte for her very competent advise on observational aspects.
  CO appreciates funding by the German Research Foundation (DFG), grant OL~350/1-1.
\end{acknowledgements}

\bibliographystyle{apj}

\bibliographystyle{aa}
\bibliography{reference}

\end{document}